# Giant anisotropic magnetoresistance in Ising superconductor-magnetic insulator tunnel junctions

Kaifei Kang[1], Shengwei Jiang[2], Helmuth Berger[3], Kenji Watanabe[4], Takashi Taniguchi[4], László Forró[3], Jie Shan[1,2,5] and Kin Fai Mak[1,2,5]

[1]School of Applied and Engineering Physics, Cornell University, Ithaca, NY, USA
[2]Laboratory of Atomic and Solid State Physics, Cornell University, Ithaca, NY, USA
[3]Institute of Condensed Matter Physics, Ecole Polytechnique Fédérale de Lausanne, Lausanne, Switzerland
[4]National Institute for Materials Science, Tsukuba, Japan
[5]Kavli Institute at Cornell for Nanoscale Science, Ithaca, NY, USA

**Superconductivity and magnetism are generally incompatible because of the opposing requirement on electron spin alignment. When combined, they produce a multitude of fascinating phenomena, including unconventional superconductivity and topological superconductivity [1-4]. The emergence of two-dimensional (2D) layered superconducting [5-13] and magnetic materials [14-18] that can form nanoscale junctions with atomically sharp interfaces [19] presents an ideal laboratory to explore new phenomena from coexisting superconductivity and magnetic ordering [20-23]. Here we report tunneling spectroscopy under an in-plane magnetic field of superconductor-ferromagnet-superconductor (S/F/S) tunnel junctions that are made of 2D Ising superconductor NbSe$_2$ and ferromagnetic insulator CrBr$_3$. We observe nearly 100% tunneling anisotropic magnetoresistance (AMR), that is, difference in tunnel resistance upon changing magnetization direction from out-of-plane to in-plane. The giant tunneling AMR is induced by superconductivity, particularly, a result of interfacial magnetic exchange coupling and spin-dependent quasiparticle scattering. We also observe an intriguing magnetic hysteresis effect in superconducting gap energy and quasiparticle scattering rate with a critical temperature that is 2 K below the superconducting transition temperature. Our study paves the path for exploring superconducting spintronics [24] and unconventional superconductivity [25] in van der Waals heterostructures.**

Two-dimensional (2D) superconductors, such as NbSe$_2$ [8], TaS$_2$ [9], and heavily doped MoS$_2$ [6, 7], are ideally suited for studies of the spin physics in superconductivity. In contrast to bulk superconductors, the orbital depairing effect is strongly suppressed under an in-plane magnetic field because the sample thickness is substantially smaller than the superconducting penetration depth [26]. These materials also possess strong Ising spin-orbit coupling that pins the electron spins in the out-of-plane direction (Fig. 1a) and results in extremely high upper critical fields on the order of 10's Teslas in the in-plane direction. At the same time, 2D crystalline insulators, such as CrI$_3$ and CrBr$_3$, possess long-range magnetic order [17, 18] with out-of-plane anisotropy. The magnetization can be aligned to the sample plane by a relatively small in-plane magnetic field (Fig. 1a). This enables us to bring to light spin-dependent effects in superconductor-ferromagnet (S/F) heterostructures. In this work, we report a new transport phenomenon in the form of giant superconductivity-induced tunnel AMR, as well as a proximity effect in the form of hysteretic quasiparticle properties under an in-plane magnetic field.

We study $NbSe_2/CrBr_3/NbSe_2$ planar tunnel junctions (Fig. 1a, b), which consist of 2-3-layer $NbSe_2$ on each side and a 4-layer $CrBr_3$ barrier. The tunnel junctions are encapsulated between hexagonal boron nitride (hBN) substrates for protection from the environmental effects. We choose 2-3-layer $NbSe_2$ samples instead of monolayers because they exhibit similar Ising superconductivity with a higher superconducting transition temperature $T_c$ (~ 6 K), and are much friendlier to work with. The thickness of $CrBr_3$ is chosen to achieve appreciable quasiparticle tunneling but negligible Josephson coupling. The Curie temperature of 2D $CrBr_3$ is reported to be around 28 K [27]. High-quality magnetic tunnel junctions with $CrBr_3$ and similar barrier materials have been recently demonstrated [28-33].

We investigate the transport properties of the junctions under a varying in-plane magnetic field using tunneling spectroscopy. The tunneling spectrum is obtained by measuring the differential conductance $G$ as a function of DC bias voltage. Multiple devices of the same structure have been studied and they all show similar behaviors. The results for device S1 are reported in the main text, and the results for the rest devices (S2-S4), in Extended Data Fig. 1. The magnetic properties of $CrBr_3$ are characterized in situ by magnetic circular dichroism (Extended Data Fig. 2). It shows out-of-plane anisotropy. At 1.6 K, the coercive field is about 10 mT and the in-plane saturation field is 0.3 - 0.4 T (the exact value is sample-dependent). Because of the relatively large magnetic anisotropy, 2D $CrBr_3$ with micron dimensions has a tendency to form single magnetic domains at low temperatures [27]. Details of device fabrication and measurements are described in Methods.

Figure 1c shows the tunneling spectra for device S1 (with 3-layer $NbSe_2$) in the absence of a magnetic field. When temperature is lowered below ~ 6 K, coherence peaks emerge at finite bias voltages. The peak separation increases and then saturates with a further decrease in temperature. At the same time, a zero-bias peak that is gradually suppressed at low temperatures is also observed. These behaviors are fully consistent with that of a superconductor-insulator-superconductor (SIS) tunnel junction [34]. The tunneling conductance corresponds to the density of states of the sample. The coherence peaks manifest the emergence of a superconducting gap. The energy at which the peaks occur is about twice of the superconducting gap. The zero-bias peak is a result of thermally activated tunneling across the gap.

The results in Fig. 1c are well described by the semiconductor model for quasiparticle tunneling (solid lines) [34]. In the model, we have used the Bardeen-Cooper-Schrieffer (BCS) density of states and a constant quasiparticle broadening parameter of $\Gamma \approx 80\ \mu eV$. Details of the analysis are described in Methods. The extracted gap $\Delta$ as a function of temperature is shown in Fig. 1d. It is in good agreement with the BCS theory with a zero-temperature gap of 0.86 meV (solid line). Moreover, the Arrhenius plot in Fig. 1e for the zero-bias conductance as a function of inverse temperature (up to 6 K) corresponds to an activation energy of 0.72 meV (dashed line). The value is in reasonable agreement with the zero-temperature gap. The temperature dependence of the zero-bias conductance in a broader temperature range (inset of Fig. 1e) also shows a dip around 28 K. It corresponds to the Curie point of CrBr$_3$ [27].

We study in Fig. 2a the evolution of tunneling spectra at the base temperature of 1.5 K when the in-plane magnetic field is increased from 0 to 1 T. The direction of magnetization in CrBr$_3$ is gradually turned from out-of-plane to in-plane above the saturation field (Fig. 1a). Changes in tunneling spectra are discernable up to ~ 0.4 T, which matches well the in-plane saturation field of CrBr$_3$. Figure 2b illustrates linecuts at representative bias voltages denoted by arrows from Fig. 2a. We observe large magnetoresistance near the coherence peak (1.3 and 1.8 mV bias). In contrast, above the coherence peak (3.0 mV) or at zero bias, tunneling conductance $G$ is nearly field-independent. The systematic bias dependence of tunneling AMR (TAMR $\equiv \frac{G_{||} - G_{\perp}}{G_{||}}$) (Fig. 2c) is evaluated using the conductance at 0 T and 1 T for $G_{||}$ and $G_{\perp}$, respectively. Giant TAMR up to ~ 100% occurs near the coherence peaks and changes sign across them. In contrast, TAMR is only 2% at large bias voltages, as well as above $T_C$ (Extended Data Fig. 3). The TAMR is therefore enhanced by nearly 50 times with the emergence of superconductivity.

The observed giant TAMR is originated from spin-dependent physics rather than from orbital effects. Large magneto-resistance is seen only below ~ 0.4 T, which is very close to the in-plane saturation field of CrBr$_3$. This saturation field is orders of magnitude smaller than the orbital-limited upper critical field ~ 200 T (estimated from the sample thickness and the in-plane coherence length ~ 10 nm) and the observed paramagnetic-limited upper critical field ~ 35 T in 2-3 layer NbSe$_2$ (Ref. [8]). Also, similar tunneling studies on non-magnetic tunnel junctions under in-plane magnetic fields do not show obvious magneto-resistance in the same range of magnetic fields (Extended Data Fig. 4).

Spin-dependent Andreev reflection has been proposed to achieve large AMR at superconductor-ferromagnetic metal junctions with spin-orbit interactions [21, 22]. A spin-flip Andreev reflection process becomes allowed when the magnetization is aligned perpendicular to the spin-orbit magnetic field, which leads to a large increase in the tunnel conductance near the coherence peaks. The process is not applicable to our SIS junctions that have no gapless quasiparticles available for Andreev reflection. Our result also shows a small spectral weight reduction (integrated between − 10 and + 10 mV) when the magnetization is aligned in-plane (Fig. 2d), i.e. perpendicular to the Ising spin-orbit magnetic field. This is in contrast to the predicted enhanced spectral weight due to spin-flip Andreev reflection [21, 22].

We perform a careful analysis of the field-dependent tunneling spectra in Fig. 2a. They show increased broadening of the coherence peaks and a small blue shift in the peak position with increasing field. This is more clearly seen in Fig. 2e and inset for the normalized tunneling spectrum at two in-plane fields of 0 and 0.5 T. We extract the values for the superconducting gap $\Delta$ and the quasiparticle broadening parameter $\Gamma$ by comparing experiment to the semiconductor model with BCS density of states (solid lines, Fig 2e) as previously performed in the absence of field. The extracted $\Delta$ and $\Gamma$ as a function of in-plane field are shown in Fig. 2f. Both $\Delta$ and $\Gamma$ increase with magnetic field between 0 T and $\sim 0.4$ T, and become nearly constant above $\sim 0.4$ T. The total change in magnitude for $\Delta$ ($\sim 20\,\mu eV$) is comparable to that for $\Gamma$ ($\sim 40\,\mu eV$). The complicated bias dependence of the TAMR (Fig. 2c) and the corresponding weak field dependence of the spectral weight (Fig. 2d) is a result of the field dependent $\Delta$ and $\Gamma$.

The increase of the superconducting gap $\Delta$ with magnetic field can be understood in terms of a magnetic exchange coupling ($\propto \boldsymbol{m} \cdot \boldsymbol{g}$) at the NbSe$_2$-CrBr$_3$ interface. Here $\boldsymbol{m}$ is the magnetization vector of CrBr$_3$ and $\boldsymbol{g}$ is the spin-orbit magnetic field of NbSe$_2$, which defines the spin quantization axis of NbSe$_2$, that is, perpendicular to the sample plane. The magnetic exchange coupling induces a Zeeman splitting in the quasiparticle density of states and is responsible for paramagnetic depairing of Cooper pairs [35]. Such paramagnetic depairing from interfacial exchange coupling is maximum when $\boldsymbol{m}$ is along the direction of $\boldsymbol{g}$ (i.e. out-of-plane). The effect vanishes when $\boldsymbol{m}$ is reoriented to the sample plane by the external magnetic field. This results in an increase of $\Delta$ (Fig. 2f), as well as the superconducting $T_C$ (Extended Data Fig. 5). Using the total increase in $\Delta$ $\sim 20\,\mu eV$ from the experiment and a hole g-factor of $\sim 10$ for TMD materials [36], we estimate the interfacial exchange magnetic field to be $\sim 0.1$ T. Zeeman splitting of the coherence peaks due to such a small interfacial magnetic exchange cannot be spectrally resolved in the tunneling spectrum at zero magnetic field (Fig. 1c).

The magnetic-field dependence of the quasiparticle broadening parameter $\Gamma$ can be explained by an interfacial spin-dependent scattering effect that is $\propto \boldsymbol{m} \times \boldsymbol{g}$. When the exchange magnetic field ($\propto \boldsymbol{m}$) is aligned perpendicular to the spin quantization axis of NbSe$_2$, that is, to the sample plane, it introduces mixing of the spin-split eigenstates in NbSe$_2$, which in turn increases both the intra- and inter-valley quasiparticle spin-flip scattering rates. Because of spin-charge separation in superconductors [24], quasiparticles near the superconducting gap edge are charge-neutral spin excitations. We can therefore estimate the spin lifetime or the quasiparticle lifetime to be $\sim 20$ ps near the gap edge from the change in $\Gamma$ ($\sim 40\,\mu eV$) under in-plane magnetic saturation. Because of the relatively short spin lifetime, nonequilibrium spin pumping and spin accumulation at NbSe$_2$ are negligible and cannot be a dominant contributor to the observed TAMR [37] (see Methods for details).

Finally, we examine the tunneling spectra under a varying in-plane magnetic field over a wider range. The field is swept from $-9$ T to 9 T in Fig. 3a and from 9 T to $-9$ T in Fig. 3b. The temperature is at 1.5 K. A hysteresis behavior is observed albeit there is no hysteresis in the CrBr$_3$ magnetization (Extended Data Fig. 2). The field dependence of the tunneling conductance at selected bias voltages is shown in Fig. 3c. Near the coherence peak, tunneling conductance shows magnetic hysteresis, whereas no hysteresis is observed at large bias voltages (also see Extended Data Fig. 3) and for the spectral weight (Fig. 2d). The hysteresis also occurs at magnetic fields substantially higher than the saturation field $\sim 0.4$ T. We extract $\Delta$ and $\Gamma$ from the tunneling spectra as a function of in-plane field (Fig. 3d). Clear hysteresis is observed in these quasiparticle properties. With increasing temperature, the hysteresis effect weakens. An example is shown in Fig. 3e for the tunneling conductance at the coherence peak. The difference between the conductance for the forward and backward field scans $\Delta G$ in Fig. 3e is computed and shown as inset of Fig. 3f. The hysteresis shrinks with increasing temperature both in $\Delta G$ amplitude and magnetic-field range. We use the maximum $\Delta G$ to locate the threshold temperature at which the magnetic hysteresis effect vanishes in Fig. 3f. The threshold temperature is ~ 4.2 K, which is ~ 2 K below the superconducting $T_C$ of NbSe$_2$.

For a ferromagnetic insulator like CrBr$_3$ with out-of-plane magnetic anisotropy subjected to an in-plane magnetic field, the in-plane sample magnetization has quadratic field dependence until its saturation at the saturation field. No magnetic hysteresis is expected for magnetic fields higher than the saturation field. This has been confirmed by magnetization measurement as a function of in-plane magnetic field (Extended Data Fig. 2). This is also fully consistent with the absence of magnetic hysteresis for tunneling conductance at bias voltages higher than the coherence peak voltage and for temperatures above $T_C$. The fact that the magnetic hysteresis occurs only near the coherence peaks shows that it is not related to the magnetic insulator but a feature of the superconductor. This conclusion is further supported by the hysteresis in $\Delta$ and $\Gamma$ (Fig. 3d). (The absence of hysteresis for the integrated spectral weight in Fig. 2d is a result of the redistribution of spectral weight in Fig. 2c.)

Multiple control experiments have been performed including the absence of hysteresis in similar S/I/S tunnel junctions with an hBN (instead of CrBr$_3$) tunnel barrier under similar experimental conditions (Extended Data Fig. 4). Based on these results, we consider vortex trapping is unlikely the origin of the observed superconducting magnetic hysteresis (see Methods for details). Could the magnetic hysteresis be of exotic origin such as ferromagnetic superconductivity? A recent theoretical study has indeed suggested that 2D NbSe$_2$ is near the onset of ferromagnetic instability [38]. The onset of magnetic hysteresis at ~ 2 K below the superconducting $T_C$ could suggest the emergence of a new superconducting state. Whether the magnetic hysteresis in NbSe$_2$-CrBr$_3$ heterostructure is of exotic origin and whether it is related to the recently observed topological superconductivity [25] deserves more in-depth investigations in the future.

**Methods**
**Fabrication of SIS tunnel junction devices**
The NbSe$_2$/CrBr$_3$/NbSe$_2$ SIS tunnel junctions were fabricated by the layer-by-layer transfer method [19]. The heterostructure was encapsulated by hBN substrates to prevent degradation of NbSe$_2$ and CrBr$_3$; both materials are air sensitive. Details of the transfer method have been reported in Ref. [19]. In short, few-layer NbSe$_2$, CrBr$_3$ and hBN flakes were exfoliated from bulk crystals onto Si/SiO$_2$ substrates. The thicknesses of the flakes were first identified by optical microscopy and verified by atomic force microscopy (AFM). The exfoliated flakes were then picked up sequentially by a polymer stamp consisting of a thin film of polycarbonate (PC) on polydimethylsiloxane (PDMS). Upon completion of the stack, the stamp was heat-released at 180°C onto another Si/SiO$_2$ substrate with pre-patterned gold electrodes. To prevent sample degradation, the exfoliation and transfer processes were performed in a nitrogen-filled glove box with O$_2$ and H$_2$O levels below 1 ppm. Multiple devices have been examined in this study and they all show similar results. Results of device S1 are included in the main text and results of S2 – S4 are summarized in Extended Data Fig. 1.

**Tunneling measurements**

Tunneling measurements were carried out in an Oxford TeslatronPT closed cycle He-4 cryostat with a base temperature of 1.5 K. In our measurement, a harmonic AC voltage $V_{ac}$ (37 Hz) with an amplitude of 40 µV, which is superimposed on a DC bias voltage, was applied to one of the NbSe$_2$ layers. The induced AC tunneling current $I_{ac}$ at the fundamental frequency was detected by a lock-in amplifier. The tunneling conductance defined as $G = \frac{I_{ac}}{V_{ac}}$ is measured as a function of the DC bias voltages. The sample plane is mounted parallel to the magnetic field direction within 1° accuracy.

In order to ensure that the small out-of-plane magnetic field component < 50 mT (up to 3 T total field) due to small angle misalignment do not affect our conclusions, we have performed tunneling measurements under an out-of-plane magnetic field up to 100 mT (Extended Data Fig. 6). The change in the tunneling spectrum is negligible compared to the in-plane magnetic field response.

**Magneto-optical measurements**

We measure the CrBr$_3$ sample magnetization by magnetic circular dichroism (MCD). Light from a diode laser with wavelength 532 nm was focused by a microscope objective onto the junction area with a diffraction-limited spot. We have limited the optical power to ~ 1 µW to reduce effects of sample heating. The incident polarization was modulated between left- and right-handed polarized by a photoelastic modulator at 50.1 kHz. The reflected light was collected by the same objective and sent to a photodiode. The AC modulated signal and the DC light power were detected by a lock-in amplifier and a multimeter, respectively. The MCD signal is defined as the ratio of the AC signal to the DC signal. The measurement was done at 1.6 K in an Attodry 2100 closed cycle He-4 cryostat equipped with an optical microscopy setup. Both the out-of-plane and in-plane magnetic field dependences were studied (Extended Data Fig. 2).

**Fitting of tunneling spectrum**

The tunneling spectra are fitted with the semiconductor model for SIS junctions [34]. The tunneling current is expressed as

$$I = \frac{G_{nn}}{e} \int_{-\infty}^{\infty} N_{1s}(E) N_{2s}(E + eV)[f(E) - f(E + eV)]dE.$$

Here $G_{nn}$ is the tunneling conductance at the normal state, $e$ is the electron charge, $N_{1s}(E)$ and $N_{2s}(E)$ are the density of states for the two superconductors at energy $E$ normalized by their respective normal state density of states, and $f(E)$ is the Fermi-Dirac distribution. We have approximated the superconducting density of states $N_{1s,2s}$ by the Dynes's formula $N_{1s,2s} = \frac{E + i\Gamma}{\sqrt{(E + i\Gamma)^2 - \Delta^2}}$ [39]. Here $\Delta$ and $\Gamma$ are the superconducting gap and the broadening parameter, respectively. To fit the tunneling spectra, the measured spectrum was first normalized by that of the same device at 6.5 K, where the NbSe$_2$ flakes are in the normal state. Only $\Delta$ and $\Gamma$ were varied as free parameters. The integration was performed in the range of $\pm 10$ mV.

**Estimates of spin lifetime and spin-pumping**

In the main text, we have observed an increase in $\Gamma$ under an in-plane magnetic field. In order to estimate the quasiparticle or spin lifetime under in-plane magnetic saturation, we first note that $\Gamma$ near zero magnetic field and at low temperature does not reflect the quasiparticle scattering rate as the parameter saturates at low temperature (Extended Data Fig. 7). Instead, it reflects the inhomogeneous distribution of the superconductor gap in $NbSe_2$ due to sample disorder. The increase in the quasiparticle scattering rate under in-plane magnetic saturation, which contributes to an increase in the measured $\Gamma$, is equal to $\sqrt{\Gamma_S^2 - \Gamma_0^2}$ ($\Gamma_S$ and $\Gamma_0$ are the measured broadening parameter under in-plane magnetic saturation and zero magnetic field, respectively). Because the spin lifetime is expected to be substantially longer for out-of-plane magnetization alignment, which does not cause spin-mixing, the quasiparticle scattering rate under in-plane magnetic saturation is $\frac{\hbar}{\tau} \sim \sqrt{\Gamma_S^2 - \Gamma_0^2}$ and $\tau \sim 20$ ps is the estimated spin lifetime. Due to inhomogeneous broadening, however, we cannot estimate the spin lifetime for out-of-plane magnetization alignment. Using the estimated spin lifetime and the typical current density $\sim 0.1$ A/cm$^2$ in our devices, we estimate a negligible spin polarization $\sim 10^{-8}$ accumulated in $NbSe_2$ due to spin-pumping.

**Excluding vortex trapping as origin of the observed magnetic hysteresis**

We discussed briefly in the main text that trapping of superconducting vortices is unlikely the origin of the observed superconducting magnetic hysteresis. Here we outline our arguments. First, because there is negligible Josephson coupling in our junctions, no Josephson vortex can be formed and no trapping of Josephson vortices is possible.

Second, because the $NbSe_2$ thickness ($\sim 1$-$2$ nm) is substantially smaller than the out-of-plane penetration depth ($\sim 200$ nm [40]), formation and trapping of in-plane vortices in the superconductor itself is strongly suppressed. The absence of in-plane vortex trapping within the $NbSe_2$ layer is further supported by the absence of magnetic hysteresis in non-magnetic $NbSe_2/hBN/NbSe_2$ SIS junctions subjected to an in-plane magnetic field (Extended Data Fig. 4).

Third, trapping of out-of-plane vortices due to non-perfect angle alignment of the magnetic field to the sample plane can also be ruled out. As discussed above, our angle alignment accuracy is within 1 degree. Since the observed magnetic hysteresis happens largely within $\pm$ 3 T in-plane magnetic field, the corresponding out-of-plane field component is within $\pm$ 50 mT. We have examined the effect of an out-of-plane magnetic field on the tunneling spectrum up to 100 mT and observed negligible effect (Extended Data Fig. 6). The density of out-of-plane vortices from angle misalignment is therefore too small to account for the observed hysteresis in $\Gamma$ in Fig. 3e.

Forth, trapping of vortices by magnetic domain walls is unlikely because the hysteresis exists way beyond the in-plane saturation field $\sim 0.4$ T. The sample magnetization is fully saturated to the sample plane in this regime and no domain wall is present. Furthermore, as shown by Ref. [27], micron-scale 2D CrBr$_3$ has the tendency to form a single magnetic domain configuration over the entire sample.

Finally, we note that the magnetic hysteresis is also unlikely originated from stray fields from the nearby environment of the device and from sample geometry. We first make sure that no bulk NbSe$_2$ flakes, which can screen out magnetic field and produce stray fields, are close to our device region. Second, multiple devices with different sample geometry and orientations to the magnetic field have been examined (Extended Data Fig. 1). All of them exhibit very similar magnetic hysteresis.

**Data availability**
The data that support the plots within this paper, and other findings of this study, are available from the corresponding authors upon reasonable request.

**Competing interests**
The authors declare no competing interests.

**Acknowledgements**
We thank Igor Mazin, Maxim Khodas, Darshana Wickramaratne, Daniel Agterberg, Sergey Frolov and Philip Kim for helpful discussions.

**Figures**

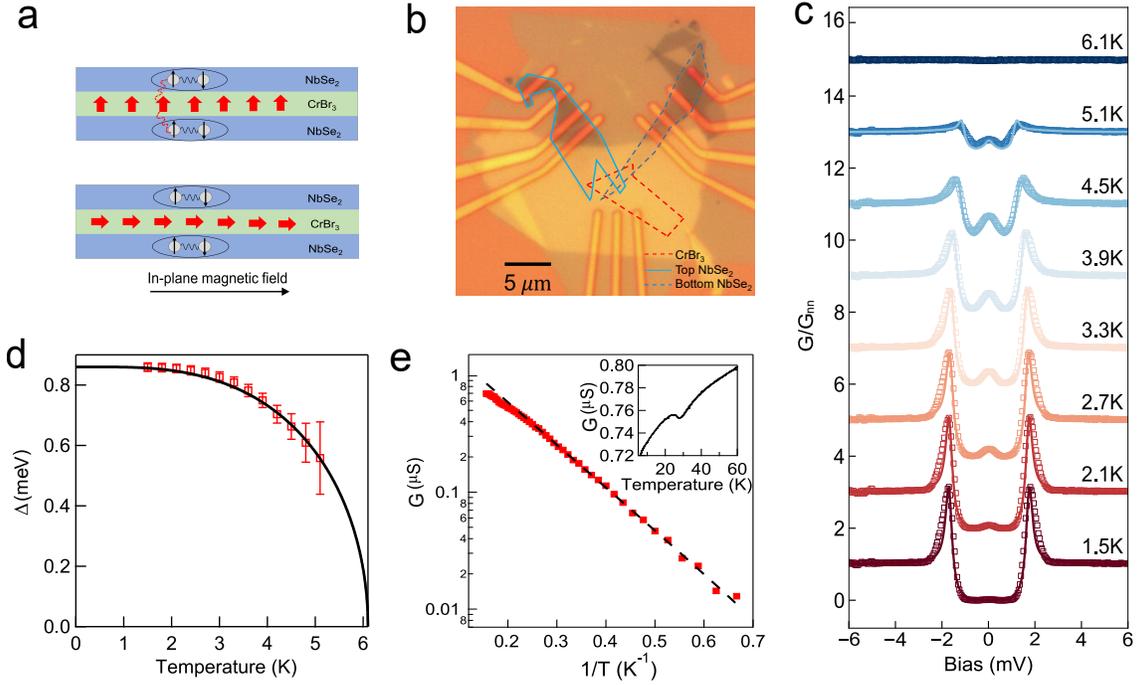

**Figure 1 | NbSe₂/CrBr₃/NbSe₂ SIS tunnel junction devices. a,** Schematics for the spin orientations (black arrows denote electron spins in Cooper pairs in NbSe₂; Red arrows denote magnetization in CrBr₃). All spins are out-of-plane (top). An in-plane magnetic field above the saturation field (∼ 0.4 T) aligns the CrBr₃ magnetization to the sample plane while the NbSe₂ spins remain largely unchanged (bottom). **b,** Optical image of Device S1. The scale bar is 5 µm. The top NbSe₂, CrBr₃ and bottom NbSe₂ flakes are marked by solid blue, dashed red and dashed blue lines, respectively. **c,** Normalized tunneling spectrum to that of the normal state at varying temperatures. Solid lines are the result of the semiconductor model described in Methods. **d,** Temperature dependence of the superconducting gap. Symbols are values extracted from the fit; error bars are the fitting uncertainty; black solid line is the BCS theory. **e,** Zero-bias tunneling conductance (symbols) below ∼ 6 K shows a thermal activation behavior with an activation gap of 0.72 meV (dashed line). The zero-bias tunneling conductance has a dip near the Curie temperature (28 K) of few-layer CrBr₃ (inset).

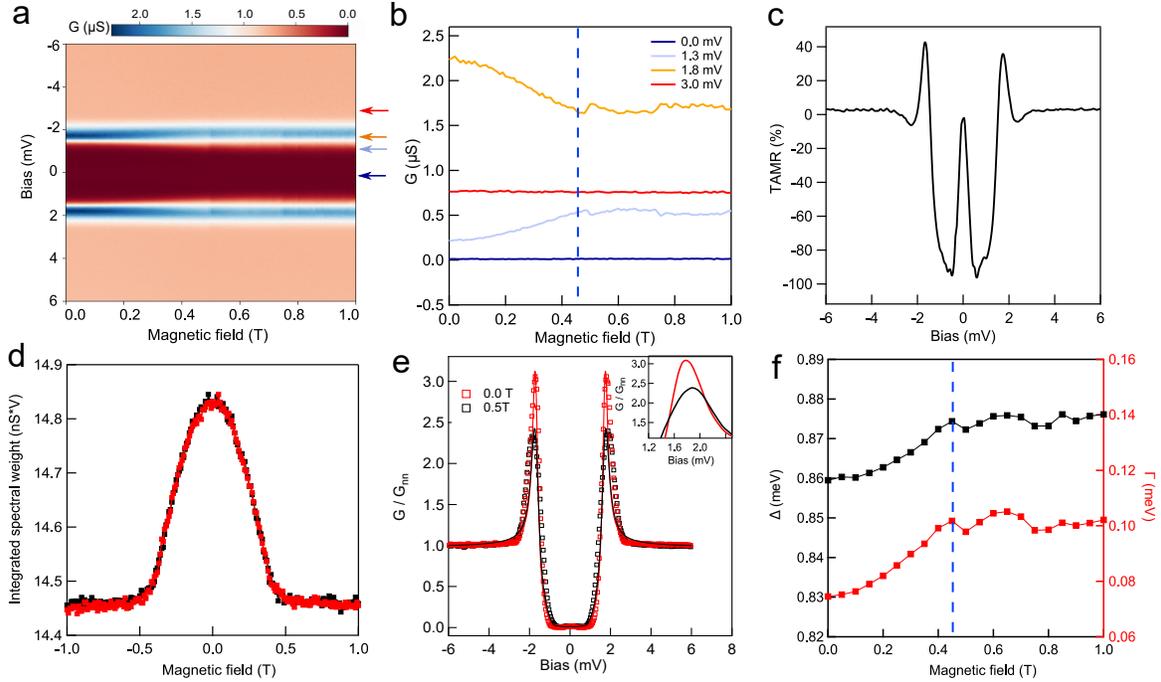

**Figure 2 | Tunneling AMR at 1.5 K. a,** Contour plot of tunneling spectrum as a function of in-plane magnetic field up to 1 T. **b,** Differential conductance as a function of magnetic field at selected DC bias voltages, 0 mV (purple), 1.3 mV (blue), 1.8 mV (yellow) and 3.0 mV (red), also marked by arrows of the same color in **a**. **c,** Bias dependence of TAMR as defined in the text. Large TAMR with sign change occurs near the coherence peaks. **d,** Integrated tunneling conductance spectral weight over a bias range of - 10 mV to 10 mV versus magnetic field. Red and Black denote two opposite field scanning directions. No hysteresis is observed. The dependence is similar to the magnetoresistance at high bias voltages or at temperatures above the superconducting $T_C$ (Extended Data Fig. 3). The result is consistent with the redistribution of spectral weight in **c**. **e,** Normalized tunneling spectrum under 0 T (red symbols) and 0.5 T (black symbols). Solid lines are the result of the semiconductor model described in Methods. The inset magnifies the coherence peak region showing a small blue shift under 0.5 T. **f,** Extracted superconducting gap Δ (black symbols) and quasiparticle broadening Γ (red symbols) as a function of magnetic field. Both increase with field and saturate above ∼ 0.45 T (dashed line). The lines are guides to the eye.

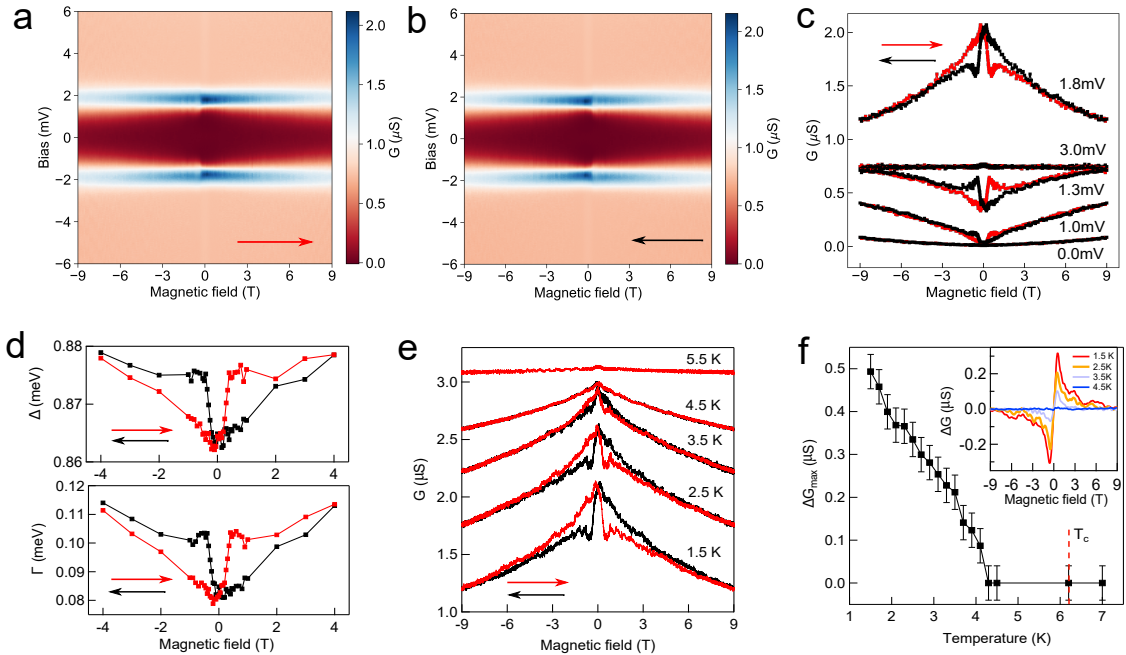

**Figure 3 | Superconducting magnetic hysteresis. a, b,** Magnetic-field dependence of tunneling spectrum in a contour plot; the in-plane field is scanned from - 9 T to 9 T (**a**) and from 9 T to - 9 T (**b**). **c,** Magnetic hysteresis of differential conductance at selected DC bias voltages at 1.5 K. **d,** Magnetic-field dependence of $\Delta$ (upper) and $\Gamma$ (lower) extracted from the tunneling spectra in **a** and **b**. **e,** Field dependence of differential conductance at the coherence peak at 1.5, 2.5, 3.5, 4.5 and 5.5 K, respectively. In **c** − **e**, black and red represent opposite scanning directions. **f,** Inset: conductance difference between the forward and backward field scanning directions calculated from the data in **e**. Magnetic hysteresis shrinks both in amplitude and field range with increasing temperature. Main: maximum conductance difference (symbol) as a function of temperature. The error bars represent the typical measurement uncertainty of differential conductance. Black line is a guide to the eye. The hysteresis disappears at ∼ 2 K below the superconducting $T_C$ ∼ 6.2 K (denoted by a red dashed line).

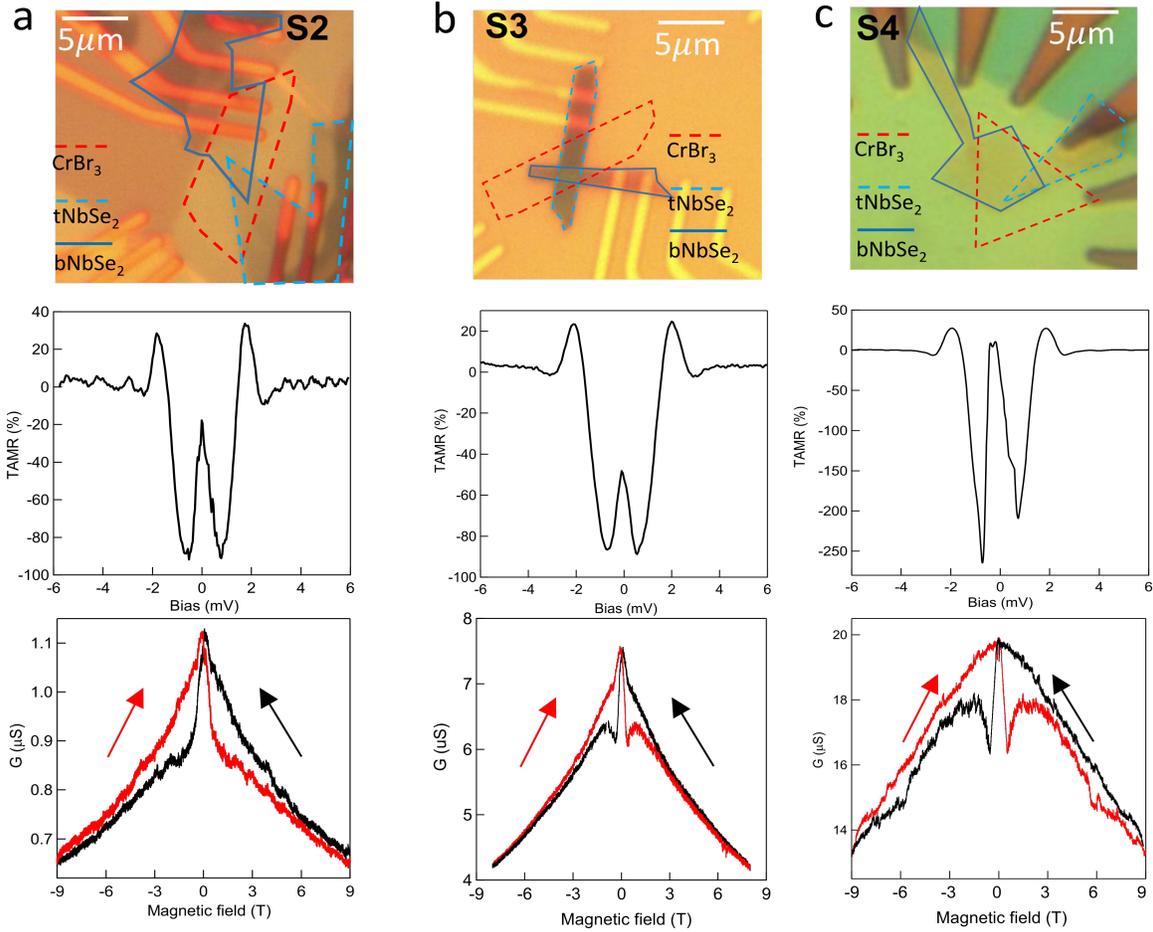

**Extended Data Figure 1 | Results from extra NbSe₂/CrBr₃/NbSe₂ junction devices. a-c,** Summary of major results from three other NbSe₂/CrBr₃/NbSe₂ magnetic SIS junctions (Device S2, S3 and S4). Top row shows the optical images of the devices. The top NbSe₂, CrBr₃ and bottom NbSe₂ flakes are marked by solid blue, dashed red and dashed blue lines, respectively. The middle row shows the bias dependence of the extracted TAMR following the same definition as in the main text. Very similar behavior is seen for all devices. The bottom row shows the in-plane magnetic field dependence of the differential conductance at the superconducting coherence peak at 1.5 K. Black and red represent reversed scanning directions. Clear magnetic hysteresis is seen for all devices, which have orientations randomly aligned to the magnetic field direction.

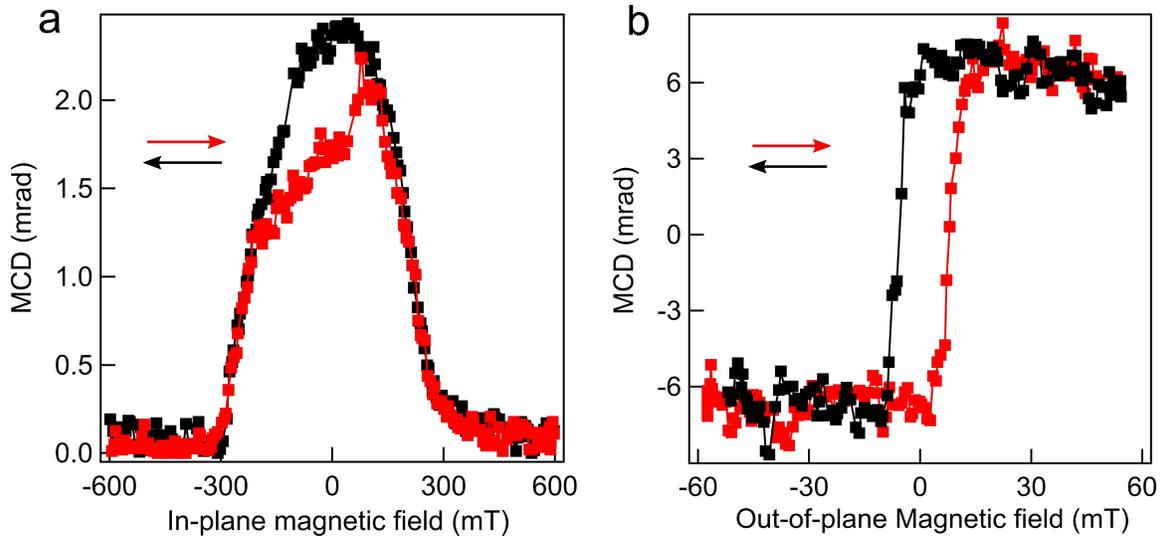

**Extended Data Figure 2 | CrBr₃ magnetization versus in-plane and out-of-plane magnetic fields.** **a,** In-plane magnetic field dependent MCD of 4-layer CrBr$_3$ (covered by 2D NbSe$_2$) under normal incidence at 1.6 K. Black and red represent reversed scanning directions. No magnetic hysteresis is seen beyond magnetic saturation. The result shows the out-of-plane anisotropy of CrBr$_3$ and the alignment of spins into the sample plane. **b,** The same for out-of-plane magnetic field dependence. A clear magnetic hysteresis is seen. The coercive field ∼ 10 mT is much smaller than the in-plane saturation field ∼ 0.3-0.4 T. The magnetic state also switches sharply at the coercive field.

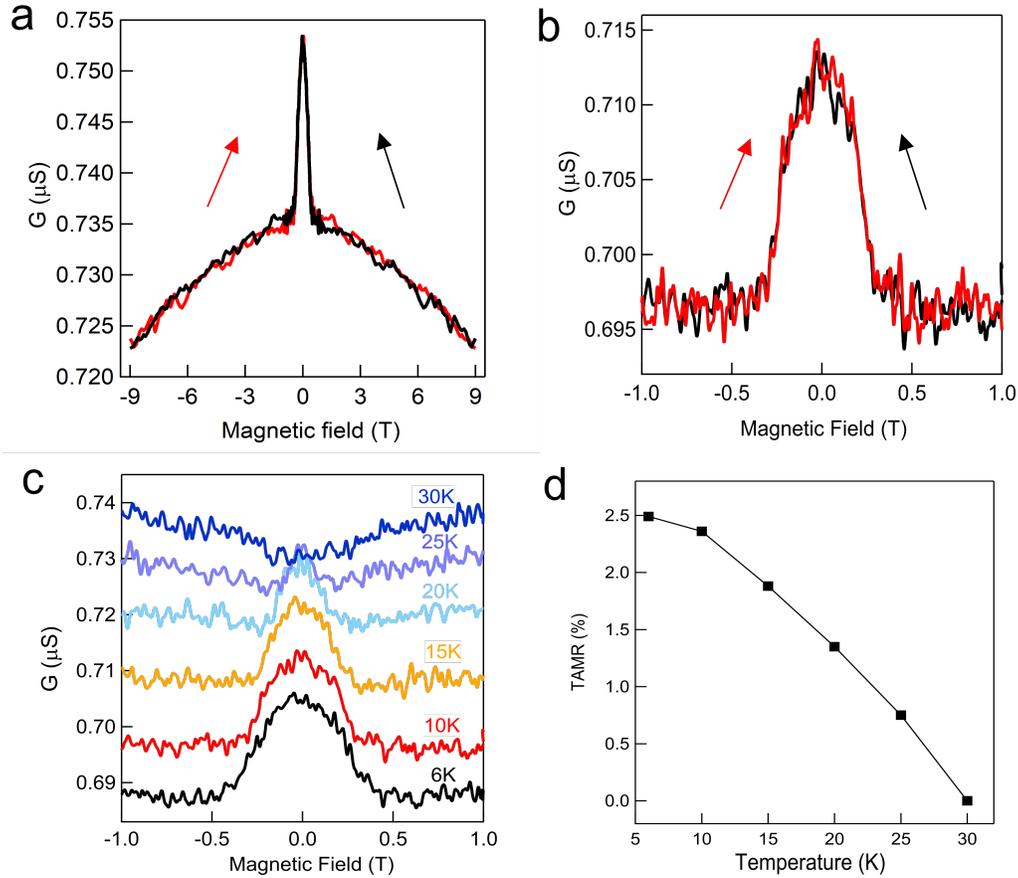

**Extended Data Figure 3 | TAMR above superconducting $T_C$ and at high bias voltages. a, b,** In-plane magnetic field dependence of the differential conductance at 5 mV bias and 1.5 K (**a**), and at 0 mV bias and 10 K (**b**). Black and red represent reversed scanning directions. No magnetic hysteresis is seen at high bias voltages and high temperatures. **c,** In-plane magnetic field dependence of the zero-bias differential conductance at varying temperatures above the superconducting $T_C$. No hysteresis is seen. The small TAMR vanishes at high temperatures. **d,** Temperature dependence of the TAMR showing the vanishing TAMR near the Curie temperature of $CrBr_3$.

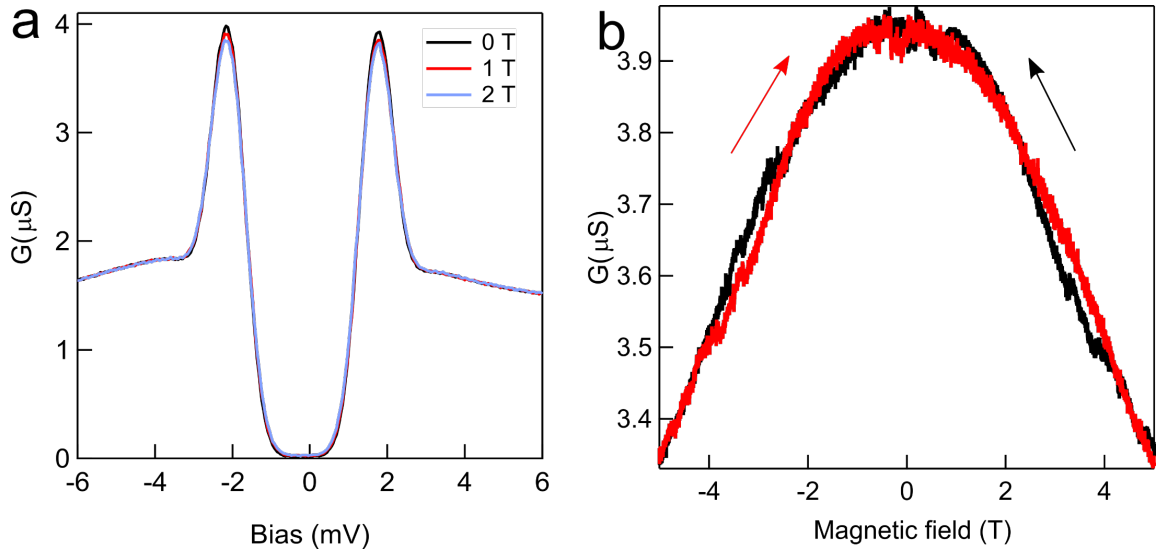

**Extended Data Figure 4 | NbSe₂/hBN/NbSe₂ non-magnetic SIS tunnel junction device. a,** Tunneling spectrum of a NbSe₂/hBN/NbSe₂ non-magnetic SIS junction at varying in-plane magnetic fields. Much weaker magnetic field dependence is seen compared to the magnetic tunnel junction. **b,** In-plane magnetic field dependence of the differential conductance at the coherence peak (1.9 mV bias). Black and red represent reversed scanning directions. No magnetic hysteresis is seen.

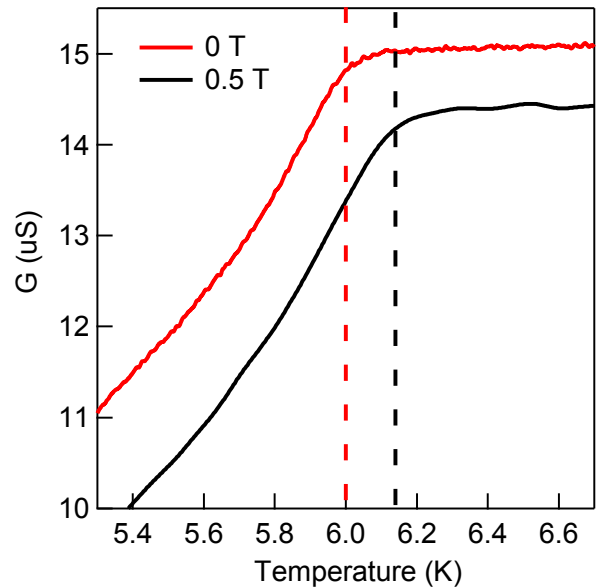

**Extended Data Figure 5 | Superconducting transition under in-plane magnetic field.** Temperature dependence of the zero-bias differential conductance of a NbSe₂/CrBr₃/NbSe₂ magnetic SIS junction at 0 T (red) and 0.5 T (black) in-plane magnetic fields. When the CrBr₃ spins are aligned to the sample plane, a small increase in the superconducting $T_C$ (the kink in the temperature dependence) by about 2 % is seen.

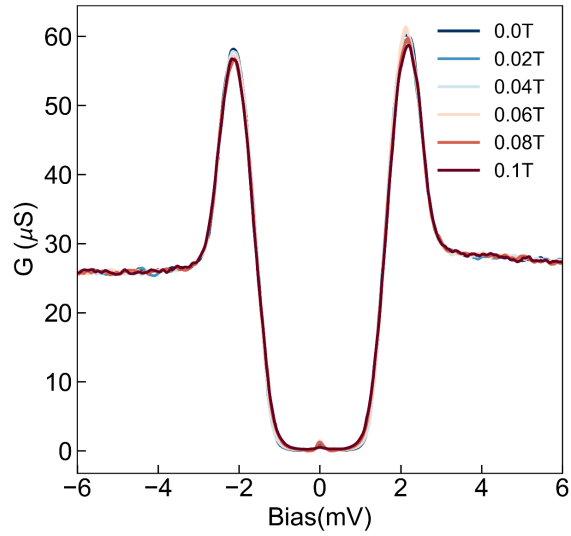

**Extended Data Figure 6 | Tunneling under out-of-plane magnetic field.** Tunneling spectrum of a NbSe$_2$/CrBr$_3$/NbSe$_2$ magnetic SIS junction at varying out-of-plane magnetic fields. No clear dependence is seen up to 100 mT. The result shows that the large in-plane magnetic field hysteresis is not the result of a small tilting of the sample plane from the in-plane magnetic field direction.

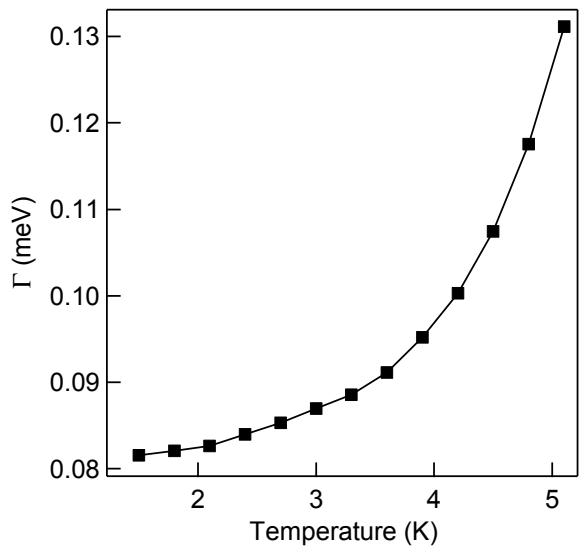

**Extended Data Figure 7 | Broadening parameter as a function of temperature.** Extracted temperature dependence of the quasiparticle broadening parameter $\Gamma$ under zero magnetic field. The increase in $\Gamma$ at high temperatures is driven by thermal broadening. The broadening saturates at low temperatures likely caused by the inhomogeneous distribution of $T_C$ in few-layer NbSe$_2$ (i.e. inhomogeneous broadening).